# Disjoint-Path Selection in Internet: What traceroutes tell us?

Sameer Qazi
Pakistan Navy Engg College (PNEC)
National Univ. Of Sciences and Technology (NUST)
Karachi, Pakistan

Tim Moors
School of Electrical Engg and Telecom
University of New South Wales (UNSW)
Sydney, Australia

***Abstract*— Routing policies used in the Internet can be restrictive, limiting communication between source-destination pairs to one path, when often better alternatives exist. To avoid route flapping, recovery mechanisms may be dampened, making adaptation slow. Unstructured overlays have been proposed to mitigate the issues of path and performance failures in the Internet by routing through an indirect-path via overlay peer(s).**

**Choosing alternate-paths in overlay networks is a challenging issue. Ensuring both availability and performance guarantees on alternate paths requires aggressive monitoring of all overlay paths using active probing; this limits scalability. An alternate technique to select an overlay-path is to bias its selection based on physical disjointness criteria to bypass the failure on the primary-path. Recently, several techniques have emerged which can optimize the selection of a disjoint-path without incurring the high costs associated with probing paths. In this paper, we show that using only commodity approaches, i.e. running infrequent traceroutes between overlay hosts, a lot of information can be revealed about the underlying physical path diversity in the overlay network which can be used to make informed-guesses for alternate-path selection. We test our approach using datasets between real-world hosts in AMP and RIPE networks.**

## I. Introduction

End-to-end paths in the Internet sometimes fail to deliver the Quality of Service (QoS) required by some applications. For many user-perceived performance failures/faults there exists an alternate path which can be used to actually prevent or "mask" the fault from the end user by using quick switch-over mechanisms. One study [1] shows that for almost 80% of the paths used in the Internet there exists an alternate path with lower probability of packet loss. On detection of failure on primary direct-path, the Internet switches to alternate direct-paths learnt through the Border Gateway Protocol (BGP). Despite being highly scalable, BGP only addresses reachability, without similar guarantees for QoS, and in the event of a failure uses a trial and error method to investigate each path in turn. When the direct-path between two hosts fails, overlay networks can quickly establish an indirect-path through intermediate host/s. It is found that the majority of direct-path failures can be bypassed by an indirect-path through a single intermediate overlay-host (*one-hop* overlay path) [2, 3](Figure 1).

Overlay-networks will only improve service if the chosen alternate-path is not affected by the failure on the primary-path. This will likely be true if the alternate-paths are disjoint from the primary. In this paper, we present an approach for selecting disjoint paths using IP-level traceroute information which can be obtained easily by running simple traceroutes between overlay peers; the idea is to select candidate indirect-paths through overlay-hosts whose paths are physically diverse from the direct-path. Furthermore, our technique does not require that the traceroutes be conducted aggressively, as this would again lead to the scalability issues, instead we propose offline processing of traceroute information. We validate our findings using real-world Internet-data from hosts in North America (AMP) [4] and Europe (RIPE*)* [5] to quantify the benefits of choosing paths that are disjoint in terms of the Autonomous Systems (ASs) they traverse.

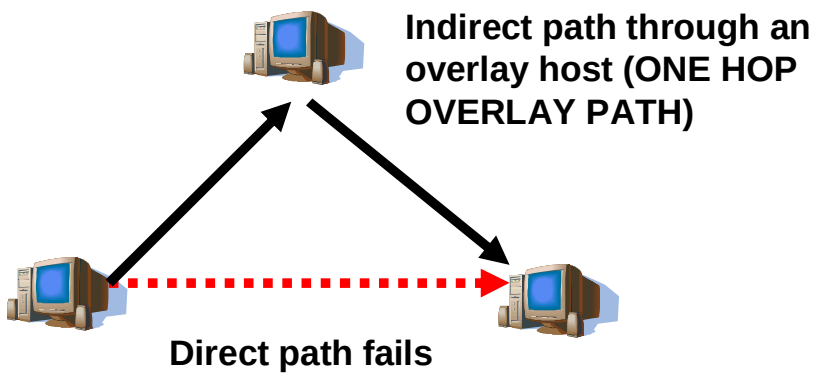

**Figure 1. Indirect-path through an overlay-host when direct-path fails.**

## II. Related Work

Finding alternate-paths to act as a backup when primary paths fail in overlay networks is previously explored in [2]. Anderson et al., designed RON to be a resilient routing tool for the Internet by implementing a small link-state overlay (50 nodes). The overlay tries to find the best alternate-path to the destination. The best path may be the default Internet path or an alternate overlay path. The design posed scalability problems due to extreme bandwidth requirements for active probing of all virtual overlay links and subsequent dissemination of this information using a link-state protocol.

Gummadi et al. [3] show that in most cases alternate paths can be found using at most one overlay hop but does not address the path selection problem. Topology aware approaches have been extensively studied to counter the scalability issue. [6] discusses a proposal for 'pruning' the overlay topology through removal of redundant physical links which are not likely to be selected by the overlay routing algorithm. Disjoint-Path selection in the Internet has been recently investigated in [7], which proposes selecting an indirect-path through an overlay node whose AS-level path digresses most-quickly from the direct-path. In our previous work [8], we proposed an approach in which we identified ASes on the most disjoint valley-free path between source-destination pair and selected overlay paths going through those ASes. We found that such approach performed better than other intuitive techniques such as the Earliest-Divergence Rule[7]. Several researchers proposed multi-path routing solutions for Internet scale networks e.g. [17-18] instead of finding explicitly disjoint paths which are sometimes not possible due to path sharing. In our recent work [19], we have shown the limitation of multi-path routing by establishing time based correlations due to difficulty in getting synchronized measurement in the network.

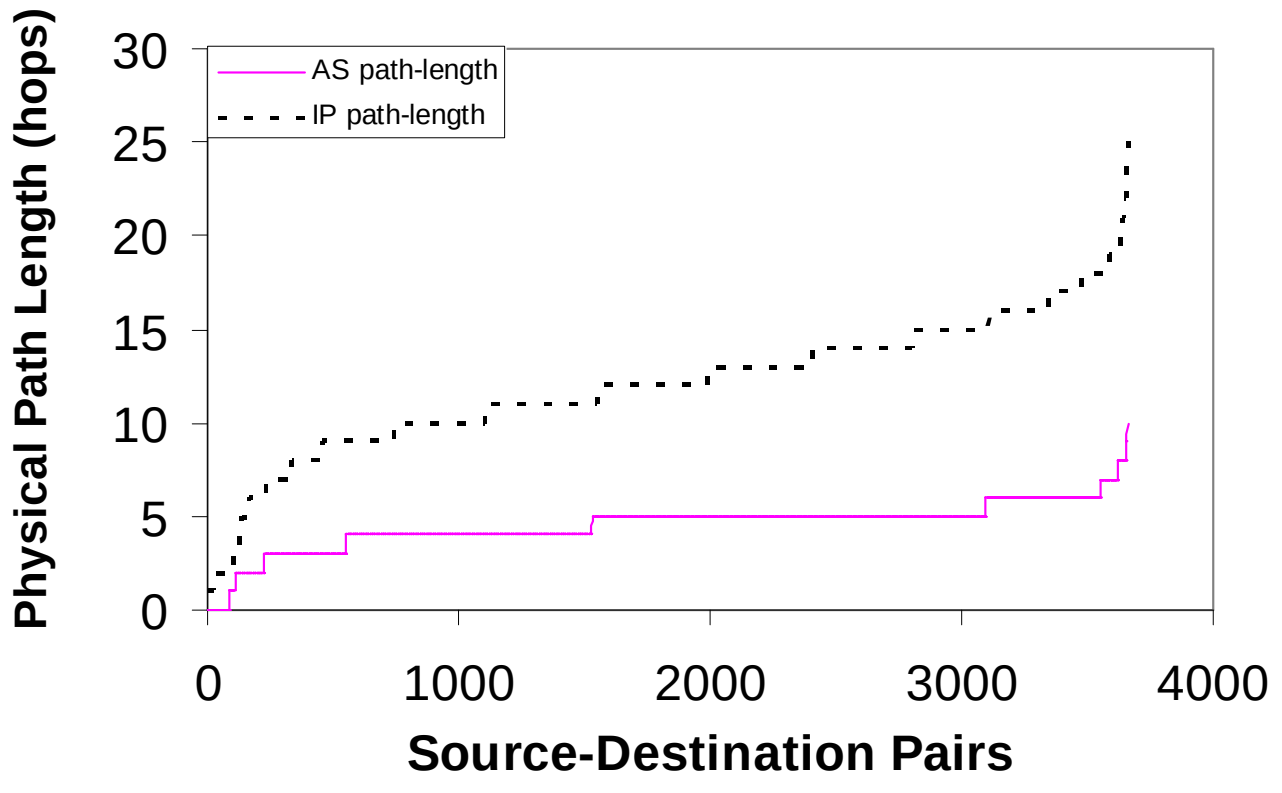

**Figure 2. Physical path length at IP level and AS level (AMP-62).**

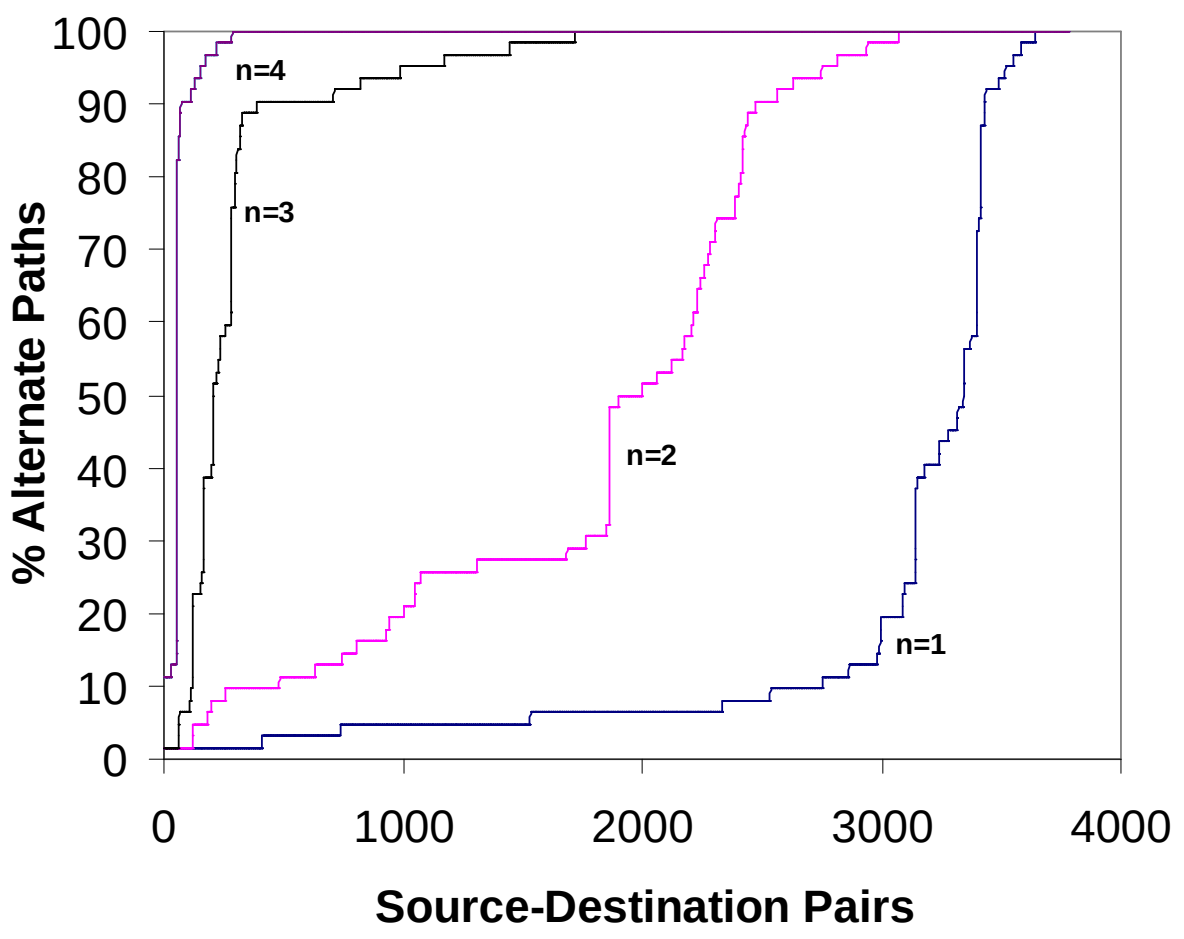

**Figure 3(a). Percentage of one-hop overlay paths which diverge at or before $n^{th}$ AS-hop with the direct path (AMP-62).**

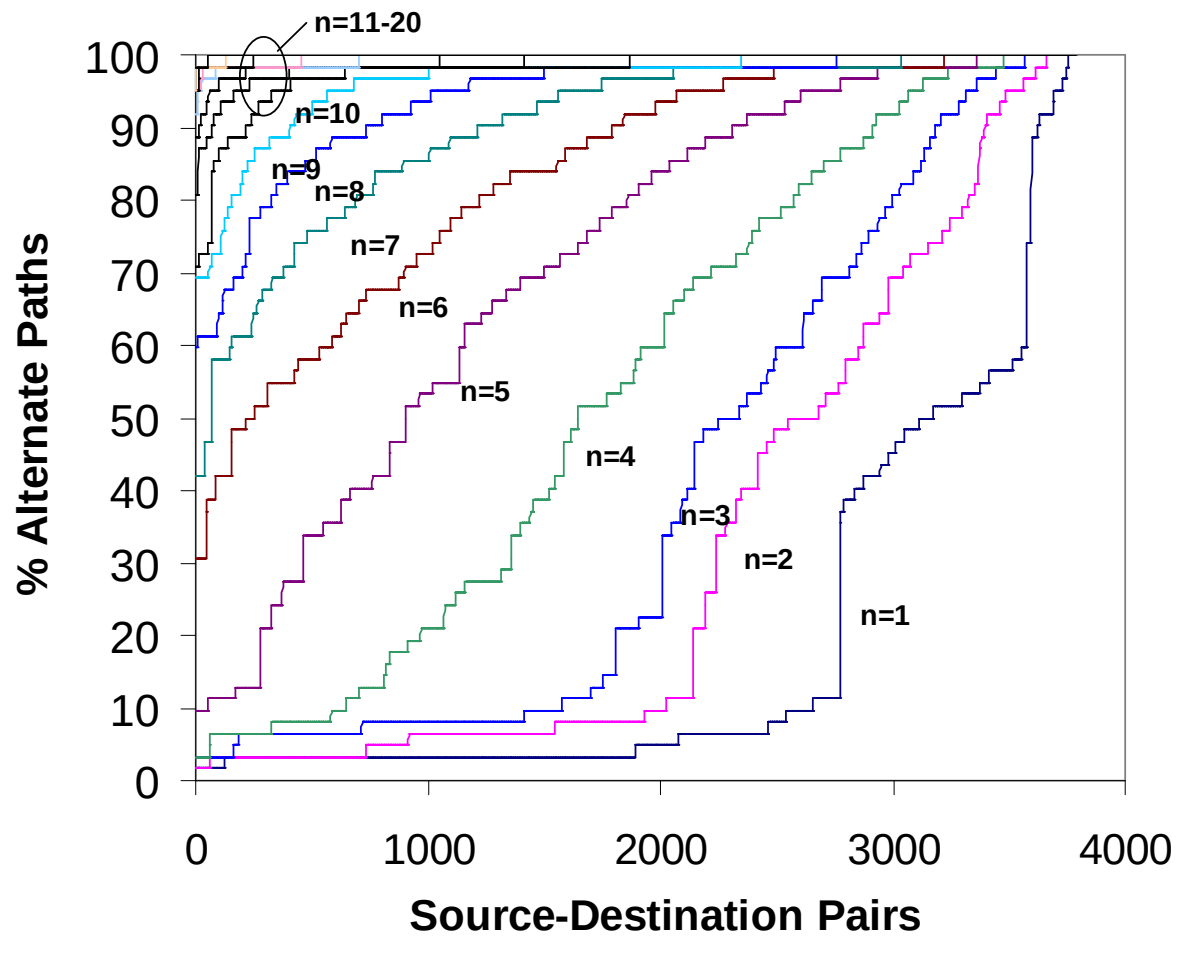

**Figure 3(b). Percentage of one-hop overlay paths which diverge at or before $n^{th}$ IP-hop with the direct path (AMP-62).**

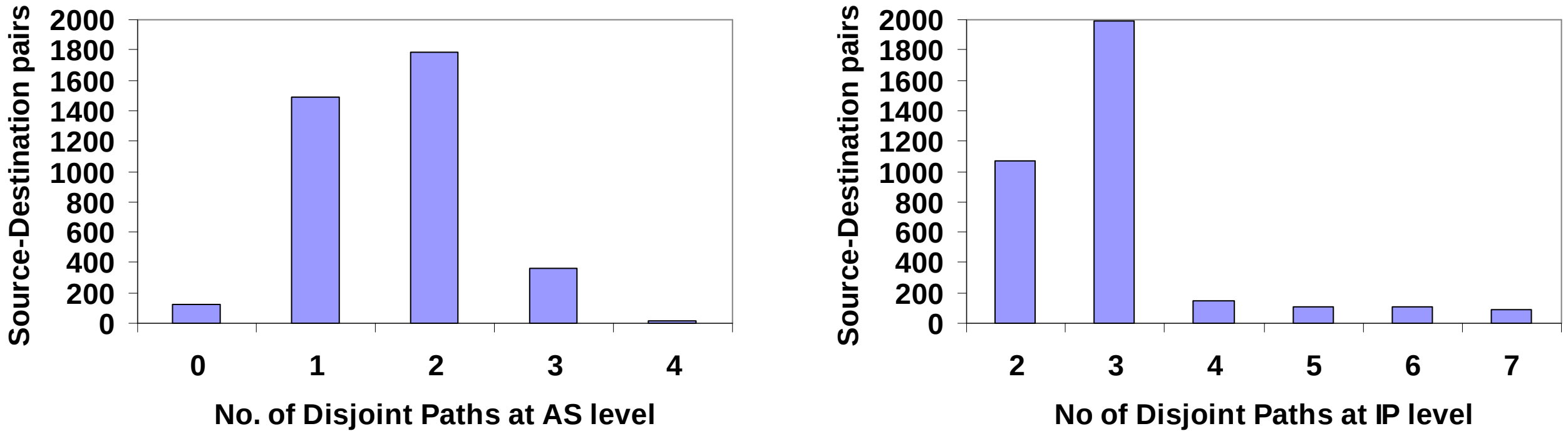

**Figure 4. Comparison of (Edge-)Disjoint paths at (a)AS level (left)and (b)IP level (AMP-62).**

## III. PATH DIVERSITY AT IP-LEVEL AND AS-LEVEL

Does IP-level knowledge really provide finer resolution of path-diversity than AS-level information? To find an answer to this question we recorded all IP-level paths using traceroutes and constructed an IP-graph for the overlay network. Each node in the graph represents a router or an end-host having a distinct IP-address and each directed-edge between them represents an IP-hop used on a valid path as observed on a traceroute.

### *Path Disjointness at IP and AS Level*

We use path and delay measurements collected between Active Measurement Project (AMP) [4] hosts from North and South America and RIPE [5] hosts, mainly from Europe and including some from Asia and North America. The datasets were collected during three 24-hr periods-June 30, 2006 (146 AMP-hosts), August 31,2006 (133-AMP-hosts) and September 5, 2007 (40 RIPE-hosts). The AMP datasets provide about one round-trip time (RTT) delay measurement for each pair of nodes per minute, and IP-traceroute information obtained around once every ten minutes. While the aim of an overlay network may only be to optimize the one-way delay, which may differ for different directions due to asymmetric Internet paths, two-way delay-measurements, such as RTTs, have been shown [9] to be strongly correlated (with a correlation-coefficient of 0.832) to one-way delays, and so form a reasonable basis for inferring one-way delays. The RIPE datasets provide two one-way delay measurements per minute and traceroutes are reported over aggregated time-intervals along with their frequency of observation. The RIPE dataset-traceroutes report both IP and AS at each hop of the path. The AMP traceroutes only report IP addresses; to map these IP addresses to AS numbers we use the IP-to-ASN Whois Service from Cymru [10], which provides mappings for user-specified dates using GNU netcat utility [11]. We refer to our virtual RON networks as AMP-XX and RIPE-XX in this paper, where XX represents the number of AMP and RIPE host selected to constiture a virtual overlay network.

Figure 2 shows that most of the AMP host-pairs have paths which traverse four Autonomous Networks or more. The corresponding length of the path in the underlying IP network is between 10 and 20 hops at an average of two to three IP hops per AS.

Figure 3 depicts the distribution of alternate paths between AMP host-pairs which diverge from the direct path at $n^{th}$ hop. Consider the percentage of alternate paths which diverge from the direct-path at AS-hop granularity (Figure 3a) and IP-hop granularity (Figure 3b). It can be seen that filtering paths based on AS-level information alone is a drastic measure as it could incorrectly categorize path offering varying degrees of IP-level path diversity as being similar in terms of path disjointness and possibly eliminate paths which can offer desirable QoS.

Next to quantify the path-diversity at the IP-level we tried to determine the number of edge-disjoint paths between AMP-hosts in the IP-graph. To determine this we again used the greedy-algorithm as in our earlier-work [7] where the algorithm starts out by eliminating the shortest-path between the AMP-host pairs in each iteration removing the edges on the path and repeating until all edge-disjoint paths have been found. However, previous studies [12] have found lack of any path diversity at the IP-layer for single-homed hosts and find that computing partially disjoint-paths where some nodes are shared amongst the paths is a better way to measure the extent of path-diversity in the IP layer. For computation of edge-disjoint paths, we take the routers on the borders of AS of the source and destination and compute disjoint-paths between them like [12]. While, at the AS-level, we find from our previous work [8] that most AMP host-pairs had one or two edge-disjoint paths, at the IP-level majority had three edge-disjoint paths and some host-pairs had up to seven edge-disjoint paths (Figure 4).

## IV. Challenges in Selecting Good Alternate IP Paths

Predicting good alternate paths that are disjoint from primary paths is easy if we know the primary path characteristics beforehand as shown in our previous work [8]. However, unlike AS paths which are stable over long periods of time, as they are dictated by commercial agreements between ISPs and routes are known publicly (through advertisements between BGP speakers); selection of paths at IP-level within ASs is challenging issue since they are changing dynamically over much smaller time-scales and are not publicly revealed. This is because within an AS, routing is dictated by internal policies of the AS inspired by the local objectives of dynamic load-balancing and hot-potato routing [13] handled by its IGP (Interior-Gateway protocol). Thus making any strong predications about the paths taken by packets through an AS virtually impossible.

Tools like traceroute which can reliably determine AS-level paths may fail to reveal the complete IP path taken by packets inside an Autonomous system This is because: routers may drop ICMP packets and further load balancing inside ASs may send traceroutes packets on a different path than the actual path taken by data packets [14]. While such tools cannot guarantee to return the path taken by the packet inside the AS it can nevertheless reveal useful information about the characteristics of the AS such as its underlying IP-layer path diversity. We follow this approach in ranking overlay nodes by the perceived IP path diversity they can offer. We find that a lot of overlay nodes which are ranked favorably by our approach are also the ones which offer the best alternate routes in the event of degradation on the direct paths between hosts.

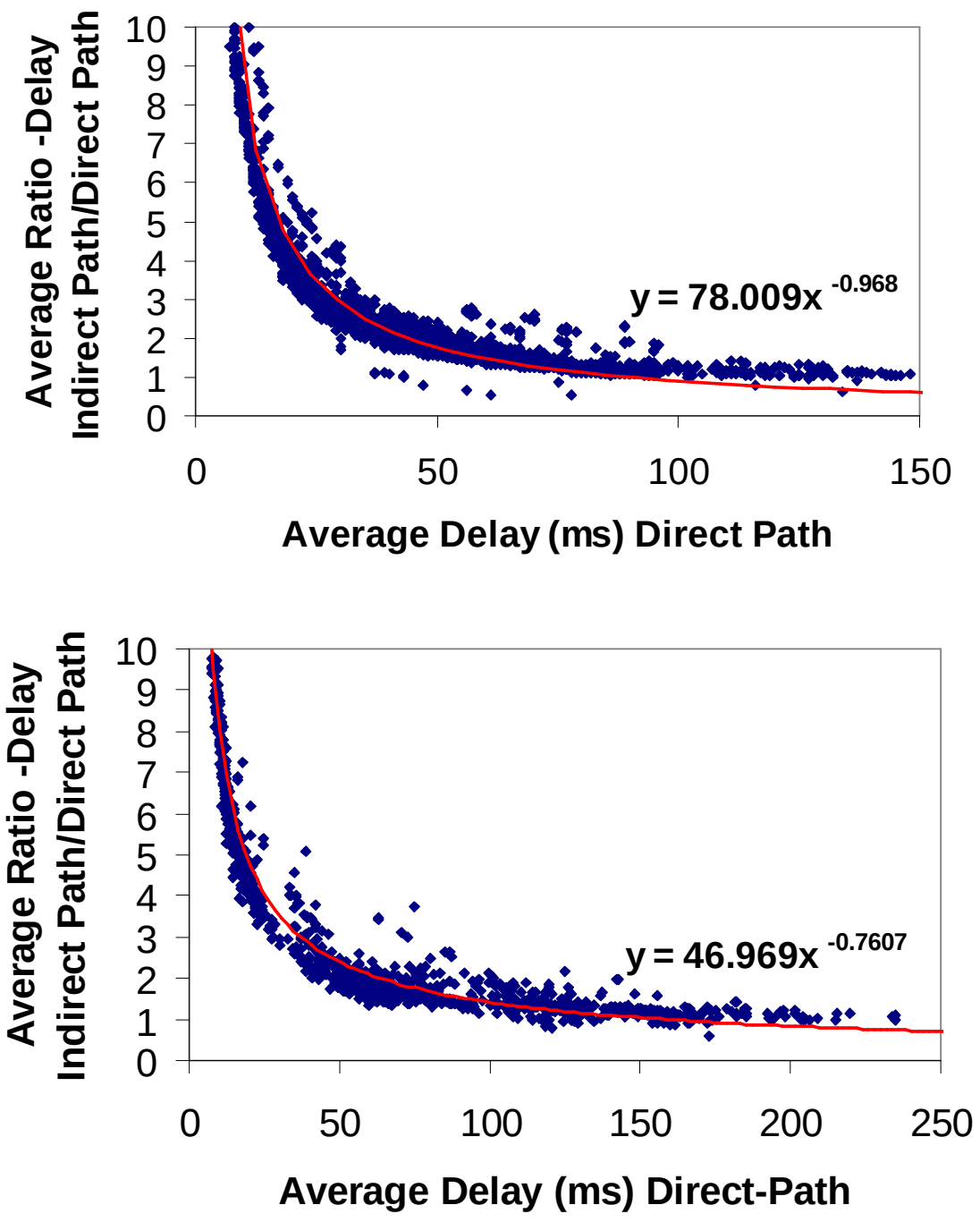

**Figure 5. Relationship of average delay on direct and indirect-paths AMP-40 (top) and RIPE-40**

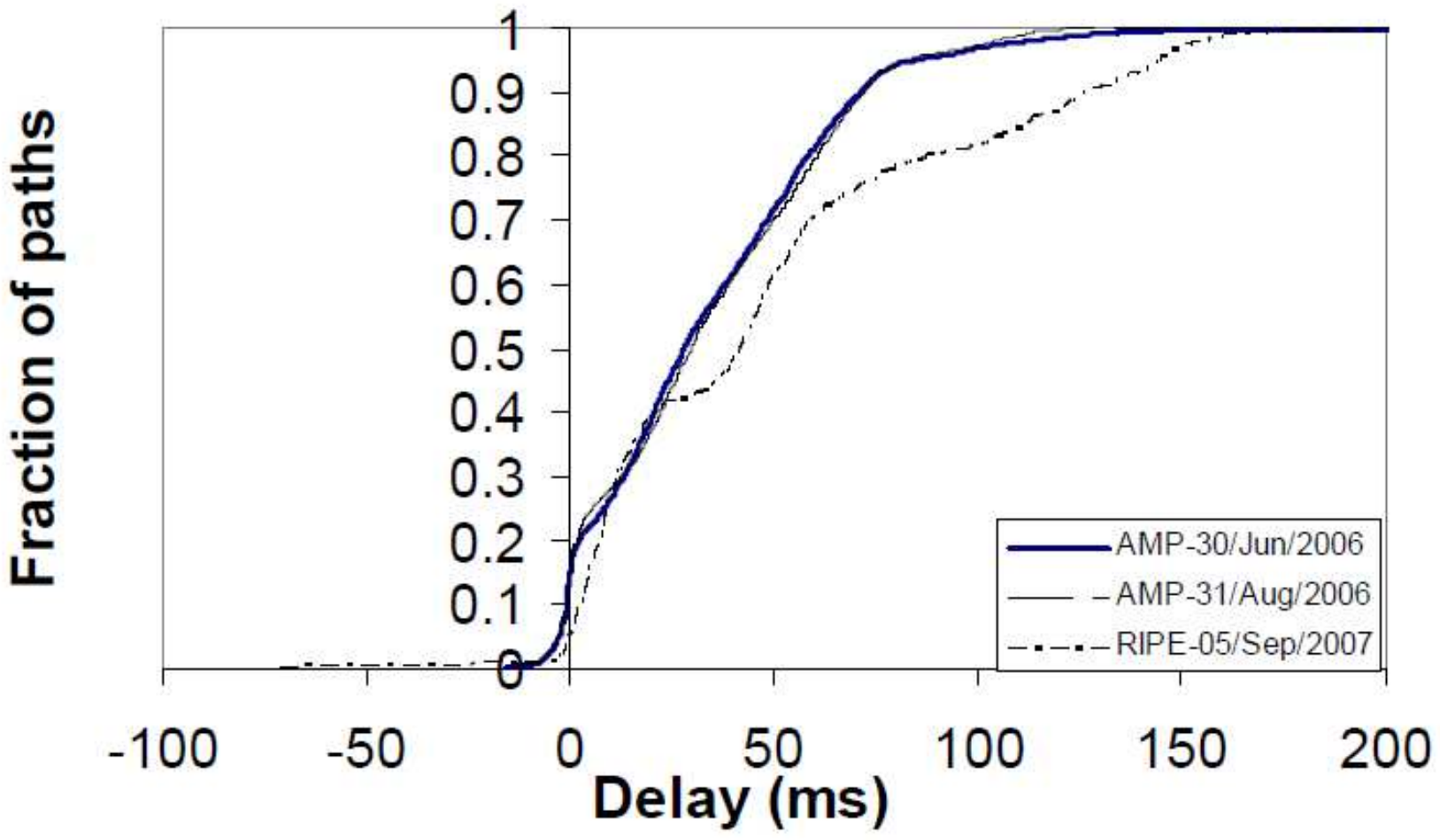

**Figure 6. CDF of the difference between mean path delay on direct internet path and the mean delay on the best one-hop overlay path.**

## V. Defining Path Degradation and Selection of Alternate Paths

The direct path between hosts in Internet is usually chosen to minimize the number of hops, which also often leads to minimizing delay. Hence, using an indirect path will usually increase delay, and so only makes sense if the current delay on the direct path is much more than the average delay on the direct path. Figure 5 shows that if the direct path has delay $D_i$, then the majority of the indirect paths have a delay ranging greater than $D_i$ according to the asymptotic relationships shown for AMP and RIPE. Figure 6 numerically quantifies the delay relationships. For 80% of the paths there is a (one-hop) alternate path providing a lower value of mean delay than the mean delay on the direct path in both RIPE and AMP networks. For AMP a majority of these alternate paths can provide up to 75 ms lower mean delay than the mean delay on the direct path. For RIPE, a majority of these alternate paths can provide up to 150ms lower mean delay than the mean delay on the direct path. The disparity in these figures is due to the fact that most of the AMP hosts are connected by high speed links on the US academic network (AMP-HPC).

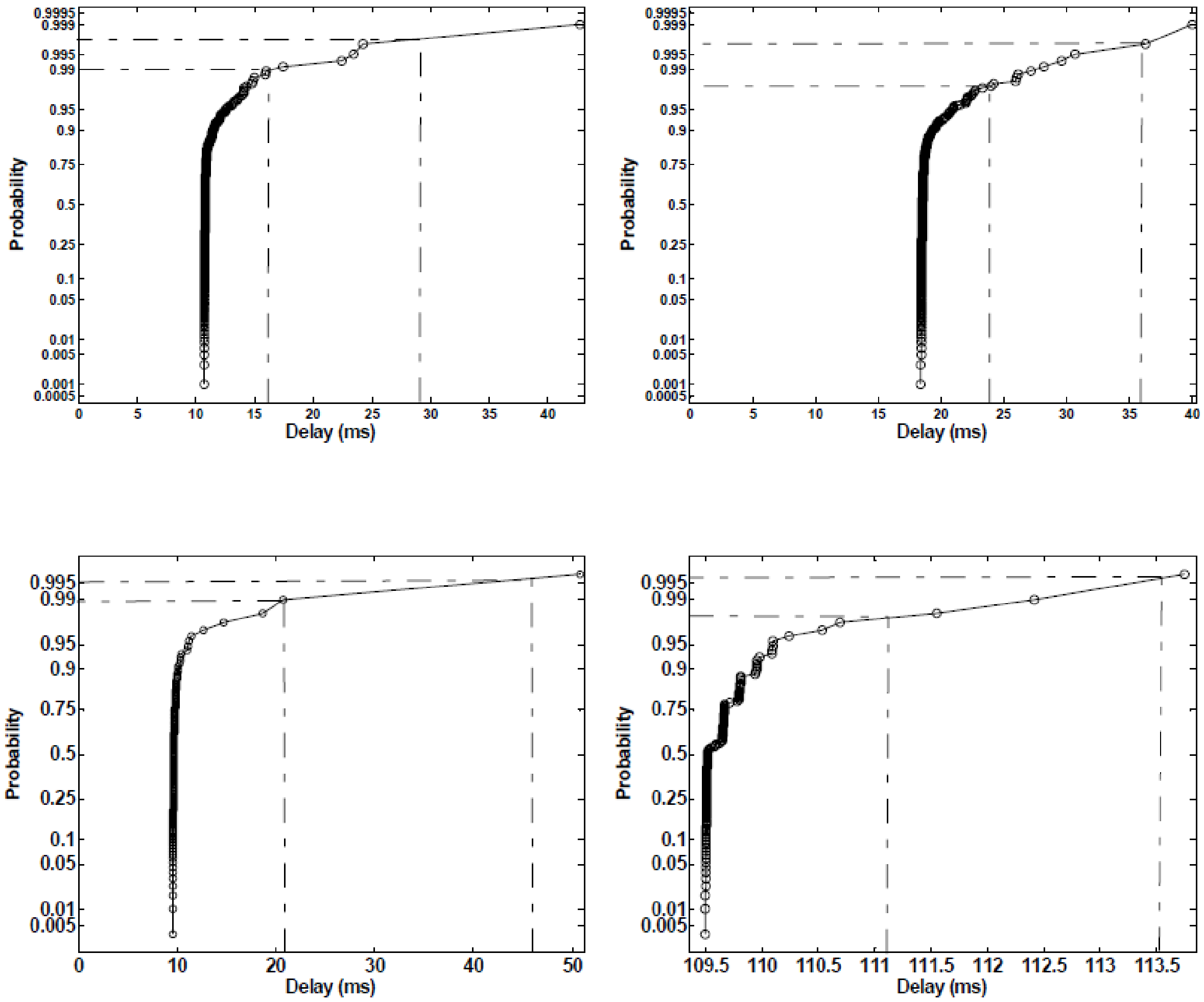

**Figure 7. Probability plots for randomly selected paths to show incidence of path outages and failures (RIPE (top) and AMP) denoted by dot-dashed lines respectively.**

We select a one hop alternate path use the same definition of a path anomaly as used by [6]. We define an anomaly as occurring when path metric (delay) exceeds its average value by a factor ( $k$ ) of the standard deviation ($\sigma$ ) of the delay values in the previous 60 epochs, one hour for AMP and 30 minutes for RIPE:

$$Path\ Delay > Path\ Delay_{average} + k\sigma$$

where $k$ =1,2,3.. is a tunable parameter to trigger an anomaly for small to large delay variations with increasing values of k , respectively. These values for k and one-hour window in determining a path anomaly are typical of those used by Chua et al. [20] and Fei et al. [6]. Chua et al. [20] worked with the Abilene network; the authors collected their network path delay measurements using NLANR AMP project measurements since a subset of AMP hosts are from the Abilene network.

Similarly, Fei et al. [6] worked with RIPE dataset. Note that the criteria for flagging a path anomaly on direct paths does not affect the relative goodness or badness of one-hop overlay paths that will be chosen to improve performance. Fei et al in [6] conjectured, "…which paths are good alternates to avoid delay degradations is relatively insensitive to the exact definition of delay degradation". In the remainder of this thesis, we refer to particular degradation considered as $k\sigma$ degradations based on the value of k used. We only select anomalies for which the immediately previous 60 epochs window do not contain any missing data. We select $k$ = 3 to emulate performance failures and $k$ =10 to emulate path outages.

Figure 7 shows probability plots for some paths on AMP and RIPE networks with thresholds for performance failures and path outages according to our definition described above for path outages and performance failures. (The averages and standard deviation are computed over the entire path delay profile). The probability of a performance failure is approximately 1-3% while the probability of a path outage is less than 0.5%.

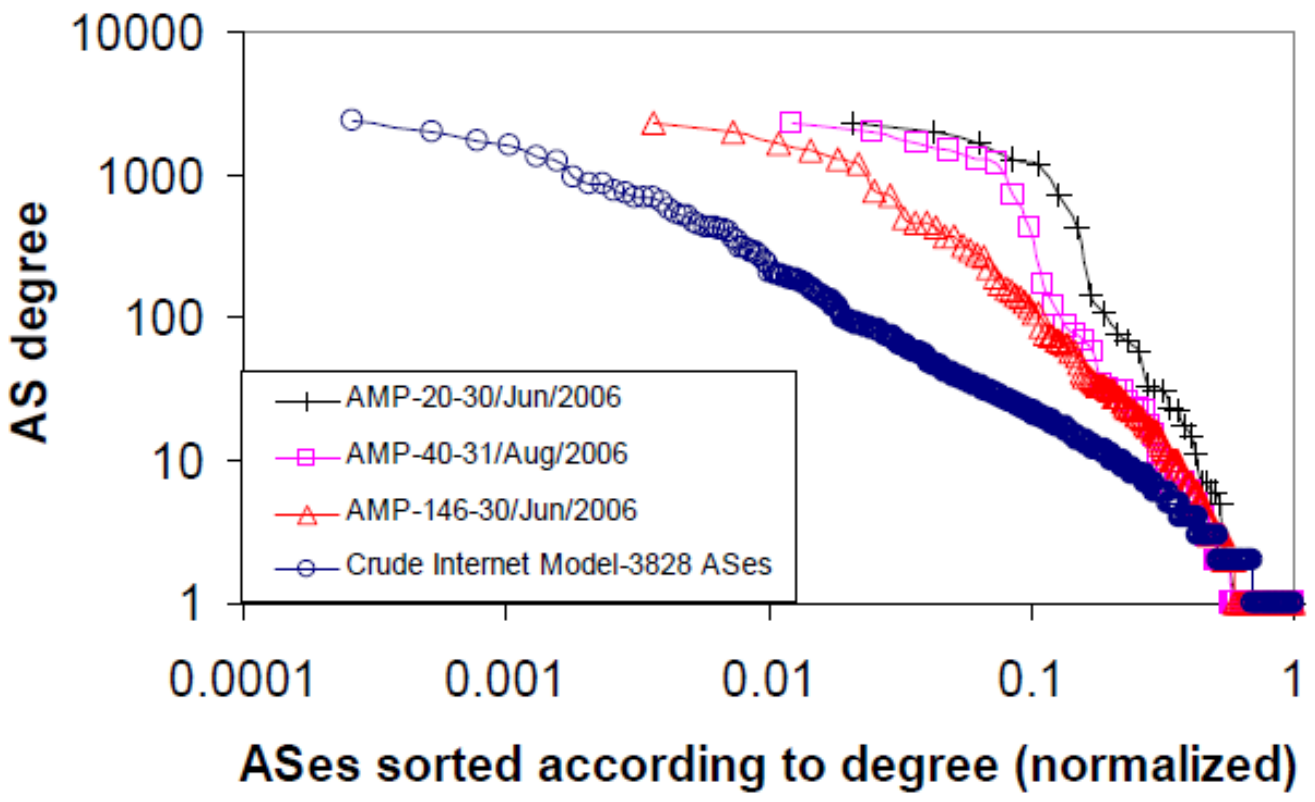

**Figure 8 . Relationship between size of an overlay network and AS degree distributions. X-axis depicts ASes sorted according to their degree-(descending order) normalized by total number of ASes.**

## VI. Relationship between Overlay Network size and path diversity it offers

The Internet topology evolves as a power-law network [23, 24]. In power-law networks, the outdegree $d_v$ of a node $v$ is proportional to the rank of the node $r_v$, to the power of a constant R i.e. $d_v \propto (r_v)^R$ [24] where $r_v$ is the index of a node in a sequence when nodes are sorted in decreasing outdegree sequence (ties in sorting are broken arbitrarily) and a typical value for $R$ is − 0.8 [24]. This means that there is a very small minority of well connected nodes which have a huge outdegree while the majority of the nodes have a very small outdegree. This power law topology phenomenon is visible in the AS level topology of the Internet; there are a few tier-1 ASes which alone constitute the majority of the inter-AS links in the Internet [24]. Customer networks are unit degree ASes (i.e. only connected to their immediate ISPs if not multi-homed) typically located at the outward fringes of the network with sparse connectivity. We next see the impact of selecting a small subset of Internet hosts for tapping into this path diversity as opposed to the billions of hosts possible. Figure 8 shows the AS degree distribution of a large number (3828) of ASes from [25] and the degree distribution of ASes sighted on overlay paths (using traceroutes) in average sized overlay networks consisting of a few tens to hundreds of AMP hosts. Notice that when even as few as 20 overlay hosts are selected to comprise an overlay network, the overlay paths already pass through the largest tier-1 AT&T network (AS 7018 with a degree of 2351). This shows that even small overlay networks can offer a substantial amount of path diversity provided the overlay hosts are in diverse ISPs to enable as much connectivity to the tier-1 & 2 networks to expose them to the AS level path redundancy in the Internet. Physically the ASes comprising the overlay network contribute to a topology that resembles a micro model of the Internet with a densely connected core and sparse connectivity at the edges. However, due to the power-law model of the Internet only a few tier-1 ASes with high connectivity are present; a majority of the customer networks are stub networks with degree of just one, i.e. only connected to their immediate ISPs which in turn rely on the large tier-1 and tier-2 ASes for connectivity to different parts (IP blocks) of the Internet. It is obvious to see as the number of hosts comprising the overlay network would increase, the network layer topology of the overlay network would tend towards the crude Internet model depicted in Figure 8. From AMP-20 to the crude Internet model, the percentage of ASes with high degree grows smaller and smaller, a reduction of two orders of magnitude in ASes with degree greater than 1000. This has the effect of stretching the graph towards the left. Due to the larger number of hosts in AMP dataset we presented the results for AMP here; RIPE would produce similar results.

## VII. Are some overlay paths preferred more often than others?

One previous study [22] has shown that some overlay paths are preferred more often than others. In their particular case, the considered overlay network was in Japan, with overlay hosts attached to geographically separated ISP's. They found that only 25% of overlay hosts were preferred more often than others, alleviating around 90% of the total failures. Similarly, Kawahara et al. [21] develop an approach for reduction in the number of transit overlay hosts based on their frequency of selection. This approach can help in selecting the optimum overlay path that provides the maximum performance benefit in a cost effective and scalable manner. We performed the same analysis on our North American and European datasets to see if this trend continued for other geographically diverse overlay networks.

Let the source node be denoted by $v_i$ and the destination node be denoted by $v_j$ ($i,j$=01,2,3,4,…$L$; $i \neq j$) where $L$ is the total number of hosts in the overlay network connected in mesh-topology. Let us define the relay nodes from $v_i$ to $v_j$ through $v_k$,($k$=0,1,2,3,…, $L$; $k \neq i,j$) at time $t$ as $v_{t,k,i,j}$ where $k$ denotes the $k^{th}$ relay node and $t$ the time at which the direct-path between $v_i$ and $v_j$ becomes degraded according to the criteria explained above. These paths are ranked by descending order of their delay gain metric as shown by (1):

$$Delaygain = \frac{D_{Direct-path} - D_{n^{th}\ Indirect-path}}{D_{Direct-path}} \qquad (1)$$

We computed the frequency with which an AMP or RIPE host in AMP-40 and RIPE-40 respectively, was amongst the top-set '*S*' [15] of a source-destination pair whose path was degraded. The top-set comprises the top *s* nodes offering the best delay-gains; we let the size of this top set to be 5, i.e. *s*=5. Let us define by *P*= (*t*,*i*,*j*), the set of those source-destination pairs ($v_i$,$v_j$) whose paths were degraded at time *t* according to our earlier definition and denote the frequency of a node being selected in the top set as $f_k$ of $v_{,k}$ (*k=0,1,2,..,L)* between $v_i$ and $v_j$ by (2) below.

$$f_k = \sum_{(t,i,j)\in P} \frac{I(v_{t,k,i,j} = v_k)}{5 * N}, (k \neq i, j) \qquad (2)$$

Where *N* is the total number of path degradations observed during the 24-hr periods the datasets were collected, it follows that |*P*|=*N.*

$$I(v_{t,k,i,j} = v_k) = \begin{cases} 1\ if \quad v_{t,k,i,j} = v_k \in S(k \neq i, j) \\ 0\ if \quad v_{t,k,i,j} = v_k \notin S(k \neq i, j) \end{cases} \qquad (3)$$

In addition, let us define the arrangement of $f_k$ in descending order of value by $f_{[z]}$ (*z*=0,1,2,3,.,*L : L=39* for AMP-40 and RIPE-40.). Then the cumulative value of $f_{[z]}$ is defined by (4).

$$F_{[z]} = \sum_{x=0}^{z} f_{[x]}, \qquad (4)$$

where $F_{[39]}$=1 holds for AMP-40 and RIPE-40 respectively.

We find that $F_{[0]}$ is about 0.1 for both AMP-40 and RIPE-40 (Figure 9). This indicates that 10% of the optimal routes can be found using only one transit node. Furthermore, 50 percent of the optimal routes can be found using 5 hosts in both AMP-40 and RIPE-40. We refer to these as the *top-5* hosts in the remainder of this paper.

## VIII. Discussion

Our finding that there exists a correlation between the size of the AS and the IP path diversity it has to offer is also intuitively logical since large ASs have many ingress and egress points into them [16] which results in good load balancing by their IGP, as

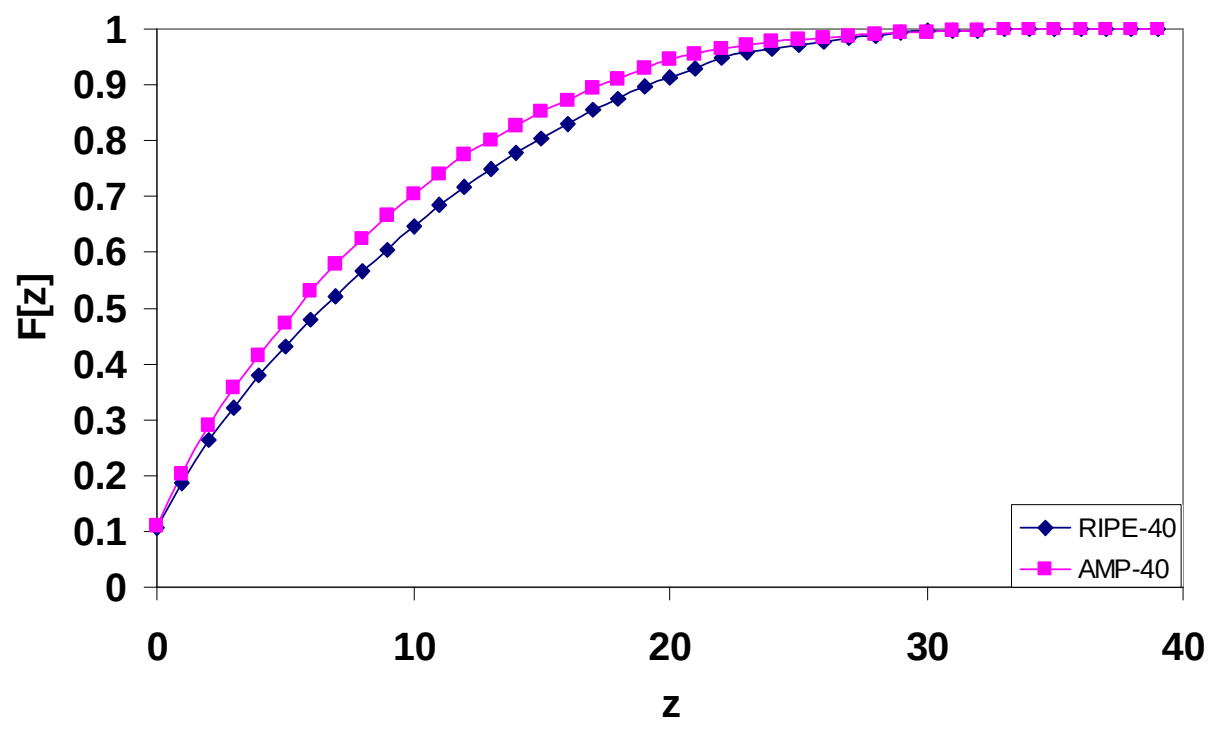

**Figure 9. Overlay hosts sorted in descending order –'z' (x-axis) according to percentage failures masked, and failures masked as Cumulative function – 'F[z]' (y-axis) for AMP-40 and RIPE-40**

well as greater IP-path diversity inside them.

The naïve approach we presented in this paper for inferring and ranking the path diversity of the ASs in order to select the best relay node, suffers from some critical issues. Routers dropping ICMP packets inside a large AS having good path diversity may erroneously cause to rank it lower in terms of path diversity amongst other ASs. Similarly, traceroutes may report multiple IP addresses corresponding to the different interfaces of the same physical router causing the AS to be ranked highly; or load balancing may send different traceroute probes on physically different paths [14] reporting incorrect information. Nevertheless, inspite of these shortcomings we see a strong correlation between the coarse path diversity inferred from only traceroutes and the probability of an overlay node in providing a QoS optimized path i.e. richness equals goodness. Improved traceroute techniques such as that specified in [14] can help to infer the path diversity of ASs more accurately and reliably.

## IX. Conclusion

In this paper we presented an approach entirely based on readily available tools such traceroutes which can be used by overlay network operators in selecting relay node offering maximum IP-path diversity for bypassing failures on direct-routes. Although arguably traceroutes suffer from several shortcomings, we find traceroutes can to some extent help in reliably determining the IP path-diversity of ASs and thus optimizing selection of a potential relay node. This would solve current scalability issues [2] associated with selection of relay nodes based on aggressive probing of the overlay network paths or computationally intensive multi-path routing approaches involving finding time based correlations between paths to rank their disjointness [18-19]. It can also supplement IP network based approaches such as IP-FRR, MPLS-FRR, MIRO and NIRA [26] for alternate path discovery.

## REFERENCES

[1] S. Savage et al., "The end-to-end effects of Internet path selection", in Proc. ACM SIGCOMM, 1999.

[2] D. Andersen et al., "Resilient overlay networks", in Proc. ACM Symposium on Operating systems principles, SOSP 2001.

[3] K. Gummadi et al., "Improving the Reliability of Internet Paths with One-hop Source Routing", In Proc. OSDI '04, 2004.

[4] Active Measurement Project (AMP). see http://watt.nlanr.net/

[5] RIPE, Test Traffic Measurements (TTM) Home Page. See http://www.ripe.net/projects/ttm/data.html

[6] A. Nakao et al., "Scalable routing overlay networks," *SIGOPS Oper. Syst. Rev.*, vol. 40(1):49-61, 2006.

[7] T. Fei et al., "How to Select a Good Alternate Path in Large Peer-to-Peer Systems?", in Proc. INFOCOM 2006

[8] S. Qazi and T. Moors, "Using Type-Of-Relationship (ToR) graphs to select Disjoint-Paths in Overlay Networks", in Proc. GLOBECOM 2007, 2007.

[9] T. Rakotoarivelo et al., "Enhancing QoS Through Alternate Path: An End-to-End Framework ", in Proc. 4th Intl. Conference on Networking (ICN), 2005.

[10] Cymru IP TO ASN Whois Service. http://www.cymru.com/

[11] GNU netcat. see http://netcat.sourceforge.net.

[12] R. Teixeira et al., "In search of path diversity in ISP networks", in Proc. IMC 2003

[13] R. Teixeira et al., "Network sensitivity to hot-potato disruptions", in Proc. SIGCOMM 2004

[14] B. Augustin et al., "Avoiding traceroute anomalies with Paris traceroute," presented at IMC '06: Proceedings of the 6th ACM SIGCOMM on Internet measurement, 2006.

[15] C.-M. Cheng et al., "Path probing relay routing for achieving high end-to-end performance", in IEEE GLOBECOM 2004.

[16] S. Zhou and R. J. Mondragon, "The rich club phenomenon in internet topology," *IEEE Communication letters*, vol. 8(3):180-82, 2004.

[17] D. Anderson et al, "Best Path Vs MultiPath Routing", in Proceedings IMC 03, pp 91-100, 2003

[18] Daria Antonova, Arvind Krishnamurthy, Zheng Ma, Ravi Sundaram, "Managing a portfolio of overlay paths", NOSSDAV 2004.

[19] S. Qazi and T. Moors, "On Issues of Multi-path Routing in Overlay - Networks Using Optimization Algorithms", Recent Progress in Data Engineering and Internet Technology, Lecture Notes in Electrical Engineering Volume 156, 2013, pp 435-440, 2013

[20] D. B. Chua, *et al.*, "Network Kriging," *Selected Areas in Communications, IEEE Journal on*, vol. 24, pp. 2263-2272, 2006.

[21] R. Kawahara, *et al.*, "On the Quality of Triangle Inequality Violation Aware Routing Overlay Architecture," in *INFOCOM 2009. The 28th Conference on Computer Communications. IEEE*, Rio de Janeiro, 2009, pp. 2761-2765.

[22] M. Uchida, *et al.*, "QoS-Aware Overlay Routing with Limited Number of Alternative Route Candidates and Its Evaluation," *IEICE Trans Commun*, vol. E89-B, pp. 2361-2374, 2006

[23] S. Zhou and R. J. Mondragon, "The rich club phenomenon in internet topology," *IEEE Communication letters*, vol. 8, pp. 180-182, March 2004.

[24] Faloutsos, *et al.*, "On power-law relationships of the Internet topology," in *SIGCOMM '99: Proceedings of the conference on Applications, technologies, architectures, and protocols for computer communication*, 1999, pp. 251-262.

[25] *CAIDA AS Relationships Dataset, see http://www.caida.org/data/active/as-relationships/.*

[26] S. Qazi and T. Moors, "Finding Alternate Paths in the Internet: A Survey of Techniques for End-to-End Path Discovery", International Journal of Current Engineering and Technology, Vol 2 No 4, PP 328-339, 2012